# The In-Flight Realtime Trigger and Localization Software of GECAM

Xiao-Yun Zhao[①], Shao-Lin Xiong*[①], Xiang-Yang Wen*[①], Xin-Qiao Li*[①], Ce Cai[①②], Shuo Xiao[①②], Qi Luo[①②], Wen-Xi Peng[①], Dong-Ya Guo[①], Zheng-Hua An[①], Ke Gong[①], Jin-Yuan Liao[①], Yan-Qiu Zhang[①②], Yue Huang[①], Lu Li[①], Xing Wen[①], Fei Zhang[①], Jing Duan[①②], Chen-Wei Wang[②], Dong-Li Shi[①④], Peng Zhang[①④], Qi-Bin Yi[①⑤], Chao-Yang Li[①⑥], Yan-Bing Xu[①], Xiao-Hua Liang[①], Ya-Qing Liu[①], Da-Li Zhang[①], Xi-Lei Sun[①], Fan Zhang[①], Gang Chen[①], Huan-Yu Wang[①], Sheng Yang[①], Xiao-Jing Liu[①], Min Gao[①], Mao-Shun Li[①], Jin-Zhou Wang[①], Xing Zhou[①②], Yi Zhao[①⑦], Wang-Chen Xue[①②], Chao Zheng[①②], Jia-Cong Liu[①②], Xing-Bo Han[③], Jin-Ling Qi[③], Jia Huang[③], Ke-Ke Zhang[③], Can Chen[①②], Xiong-Tao Yang[①], Dong-Jie Hou[①], Yu-Sa Wang[①], Rui Qiao[①], Xiang Ma[①], Xiao-Bo Li[①], Ping Wang[①], Xin-Ying Song[①], Li-Ming Song[①], Shi-Jie Zheng[①], Bing Li[①], Hong-Mei Zhang[①], Yue Zhu[①], Wei Chen[①], Jian-Jian He[①], Zhen Zhang[①], Jin Hou[④], Hong-Jun Wang[④], Yan-Chao Hao[④], Xiang-Yu Wang[④], Zong-Yuan Yang[④], Zhi-Long Wen[④], Zhi Chang[①], Yuan-Yuan Du[①], Rui Gao[①], Xiao-Fei Lan[⑥], Yan-Guo Li[①], Gang Li[①], Xu-Fang Li[①], Fang-Jun Lu[①], Hong Lu[①], Bin Meng[①], Feng Shi[①], Hui Wang[①], Hui-Zhen Wang[①], Yu-Peng Xu[①], Jia-Wei Yang[①], Xue-Juan Yang[⑤], Shuang-Nan Zhang[①], Chao-Yue Zhang[⑧], Cheng-Mo Zhang[①], Zhi-Cheng Tang[①], Cheng Cheng[⑨]

① Key Laboratory of Particle Astrophysics, Institute of High Energy Physics, Chinese Academy of Sciences, Beijing 100049, China
② University of Chinese Academy of Sciences, Chinese Academy of Sciences, Beijing 100049, China
③ Innovation Academy for Microsatellites of CAS, Shanghai 201203, China
④ School of Information Science and Technology, Southwest Jiaotong University, Chengdu 611756, China
⑤ Key Laboratory of Stellar and Interstellar Physics and Department of Physics, Xiangtan University, 411105 Xiangtan, Hunan Province, China
⑥ Physics and Space Science College, China West Normal University, Nanchong 637002, China
⑦ Department of Astronomy, Beijing Normal University, Beijing 100875, China
⑧ Changchun University of Science and Technology, Changchun 130022, Jilin, China
⑨ Chinese Academy of Sciences South America Center for Astronomy, National Astronomical Observatories, CAS, Beijing 100101, China

Contact, E-mail: xiongsl@ihep.ac.cn, wenxy@ihep.ac.cn, lixq@ihep.ac.cn

**Abstract** Realtime trigger and localization of bursts are the key functions of GECAM, which is an all-sky gamma-ray monitor launched in Dec 10, 2020. We developed a multifunctional trigger and localization software operating on the CPU of the GECAM electronic box (EBOX). This onboard software has the following features: high trigger efficiency for real celestial bursts with a suppression of false triggers caused by charged particle bursts and background fluctuation, dedicated localization algorithm optimized for short and long bursts respetively, short time latency of the trigger information which is downlinked throught the BeiDou satellite navigation System (BDS). This paper presents the detailed design and deveopment of this trigger and localization software system of GECAM, including the main functions, general design, workflow and algorithms, as well as the verification and demonstration of this software, including the on-ground trigger tests with simulated gamma-ray bursts made by a dedicated X-ray tube and the in-flight performance to real gamma-ray bursts and magnetar bursts.

**Key Words**: GECAM, trigger, localization, classification, BeiDou short messages

## 1 Introduction

Gravitational wave high-energy Electromagnetic Counterpart All-sky Monitor (GECAM) is a constellation of twin microsatellates [1] designed for the all-sky monitoring of gamma-ray transients such as gamma-ray burst (GRB), especially those associated with gravitational wave events (GW) and high energy neutrinos (HEN), magnetar bursts (SGR), X-ray binaries, solar flares (SFL) and terrestrial gamma-ray flashes (TGF).

Both GECAM satellites were launched to the same orbit with an inclination of about 29 degree and a height of 600 km in Dec 10, 2020 (Beijing time) [2]. There is no direct communication between these two satellites on orbit. As shown in Figure 1, the payload of each GECAM satellite consists of a detector dome and an electronics box (EBOX). Each GECAM satellite can monitor bursts in all sky region unocculted by the Earth in the energy range of about 10 keV to 5 MeV, and trigger and locate the burst in real time, and then downlink the trigger information to ground using the short message service of the BeiDou satellite navigation System (BDS) to facilitate the joint and follow-up observations with multi-messenger and multi-wavelength telescopes [3].

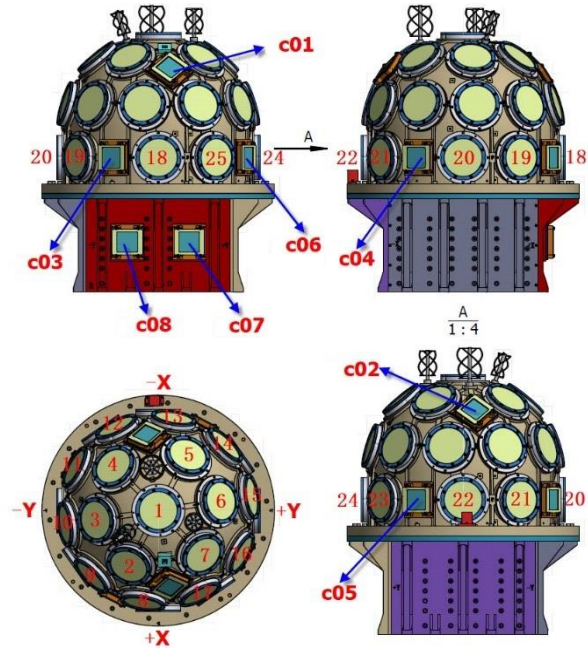

Figure 1. Schematic diagram of the GECAM payload which consists of a detector dome and an electronic box (EBOX). Different views of the payload are shown together with the payload coordinate system illustrated in the bottom left panel. 25 GRDs are shown with numbers from 1 to 25, while CPDs with labels starting with 'c'. Two CPDs are installed in the +X direction of the EBOX.

Each GECAM satellite is equipped with 25 Gamma-Ray Detectors (GRDs) and 8 Charged Particle Detectors (CPDs). All 25 GRDs and 6 CPDs are assembled in the detector dome of the satellite with different orientations to cover most of the sky region, while 2 CPDs are installed in the +X direction of the payload electronics box (EBOX) (Fig. 1). EBOX consists of one power supply board, five data acquisition boards and one data management board with co-running embedded software, thus it can supply secondary power from the spacecraft to the payload, process sicence and engineering data from detectors, handle command and data communication between payload and spacecraft [1].

Realtime trigger and localization of bursts (mostly GRBs and SGRs) is critically important for the multi-wavelength and multi-messenger observations. However, among the current gamma-ray burst missions, only Fermi/GBM [4], Swift/BAT [5] and GECAM can provide realtime trigger and localization (the location accuracy is about several degree for Fermi/GBM and 3-4 arcmins for Swift/BAT). INTEGRAL/SPI-ACS [6] and CALET/GBM [8] can provide realtime trigger but without localization, while Konus-Wind [7], Insight-HXMT/HE [9] and AstroSAT/CZTI[2] can provide trigger and coarse localization but with time latency of hours to days. The IPN [3] composed of basically all GRB missions can usually offer good localization but with significant time latency up to several days. The upcoming missions such as SVOM/ECLAIRs [10], SVOM/GRM [11] and EP/WXT [12] are also designed to have the realtime trigger and location capability.

As the first Chinese space telescope with realtime downlink capability, GECAM is designed to provid the trigger and location of high energy bursts in real time. We designed and developed a dedicated onboard trigger and localization software for GECAM which is configured and executed on the ARM Microcontroller Unit (SAMV71Q21RT-DHB-SV) of GECAM EBOX.

In this paper, we present the detailed design and deveopment of this software system.

## 2 Software Design

**2.1 Requirements**

According to the scienctific requirments of GECAM and lessons learn from previous GRB missions, we defined the following requirements for this in-flight software:

(1) Continuously evaluate triggers in multiple time scales and energy channel ranges. Calculate localization for successful triggers based on the signals of GRD detectors.

(2) Classify triggers to type I (important triggers, e.g., gamma bursts, magnetar bursts) and type II (other triggers, e.g. particles events), and generate alert data (i.e. the BDS short message) for type I triggers and engineering data (without real-time downlink) for type II triggers.

(3) Generate high time resolution light curves for short trigger (the main candidate for gravitational wave electromagnetic counterpart), which is necessary for the on-ground time-delay localization with multiple spacecrafts. For long triggers, generate ~300 s of light curves and the counts spectrum, which could be used in the on-ground refined localization.

(4) The whole software together with the settings and data tables (trigger algorithm parameter tables, localization templates, etc.) could be updated in-flight.

**2.2 Overall design**

The trigger and localizaiton software is a part of Data Management Unit (DMU) software, and they can be refreshed and re-configured in-flight.

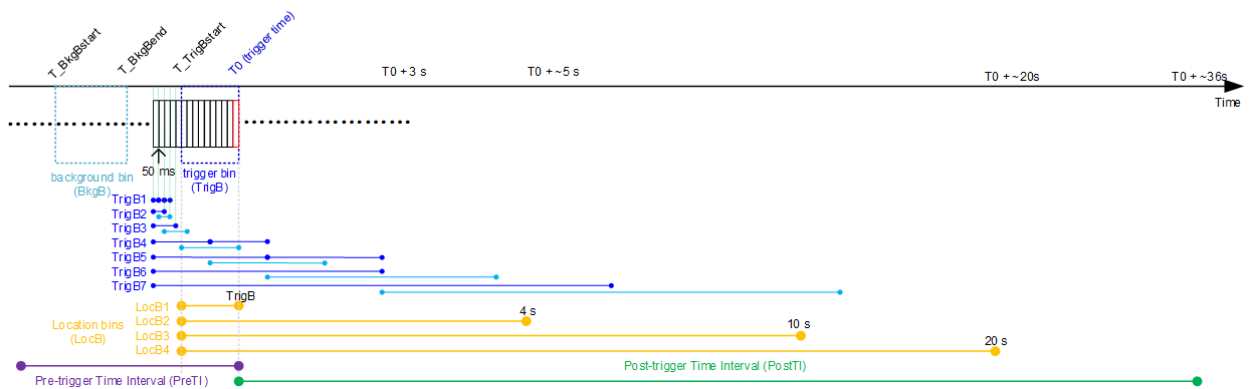

Figure 2. Illustration of the trigger and localization procedures of the GECAM onboard trigger and localization software.

The overall design of this software is shown in Figure 2. Firstly, it read the 50 ms time-binned data from the 8 channels of the 25 GRD detectors and then accumulates the background from T_BkgBstart to T_BkgBend (see Figure 2). The trigger criterion is that there are a given number of detectors having significant excess counts above the background. The trigger threshold values (e.g. the minimum number of detectors with excess counts, significance level of the counts excess, the maximum angles between triggered detectors) are read from the triger parameter table. For a successful trigger, trigger information (such as the trigger time (T0), trigger time scale and energy range, the ID of three detectors with highest significance, etc.) will be generated and stored in the first two BDS short messages (see Table 20).

The burst location will be computed in four preset timescales using data of 25 GRDs, and the localition with the best goodness of fit is chosen to be the final in-flight location. The best location is transfered to celestial coordinate system (J2000) from the payload coordiates.

For each trigger, the software will determine whether it is type I or type II trigger based on a series of information such as the trigger parameters, location of the burst and the counts ratio between CPDs and GRDs. Events that classified as particle events, solar flares, or Earth occultaion of known sources will be flaged as Type II triggers, while others are classified as Type I triggers.

It will generate alert data (i.e. the BDS short message to be downlined in near real time) for type I triggers and engineering data (without real-time downlink) for type II triggers. The software will also tell whether is is a short burst or long burst according to the light curves between T0+2 s to T0+4 s. For Type I and short triggers, the high time-resolution light curve (0.4 ms by default) of 25 GRDs within about 1 s around T0 will be generated. These data will be compressed since the number of short messages for each trigger is limited (by default 31 message for each trigger). For Type I and long triggers, it will generate ~300 s of light curves and the counts spectrum for 25 GRDs and CPDs.

## 2.3 Module design

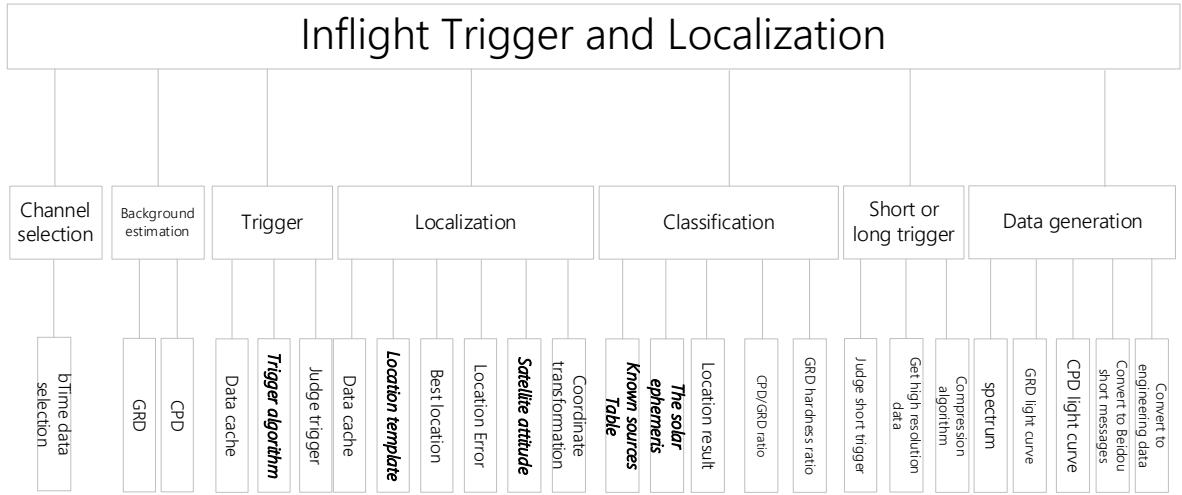

Figure 3. Functional modules of the GECAM onboard trigger and localization software.

Figure 3 shows the modules for this software. The functions of each module are briefly described as below and a further detailed implementation of each module will be introduced in section 3.

**Energy Channel Selection Module**: The input data for this software is the time-rebinned data (called BTIME data) for 25 GRDs and 8 CPDs every 50 ms. Data in the specified energy ranges will be selected according to the software setting.

**Background Estimation Module**: Provide the background estimation for both GRDs and CPDs.

**Trigger Algorithm Module**: Search for triggers in real time.

**Positioning Module:** Calculate sky location of burst (J2000 coordinates).

**Trigger Type Classification Module**: Classify the trigger to type I (important events) or type II (others).

**Short/long Trigger judement Module**: Judge the duration of a trigger and classify to short or long trigger.

**Data Generation Module**: Generate the BDS short messages for type I triggers, and the engineering data for type II triggers.

## 2.4 The workflow

The general workflow of this software is shown in Figure 4. There are 17 kinds of processes in total, and the most common processes are: the type I long trigger process, type I short trigger process, and type I short and long trigger process, which will generate 31, no more than 31 and no more than 62 short messages respectively. The number of BDS short messages for short triggers depends on compression results of the high time resolution light curve data.

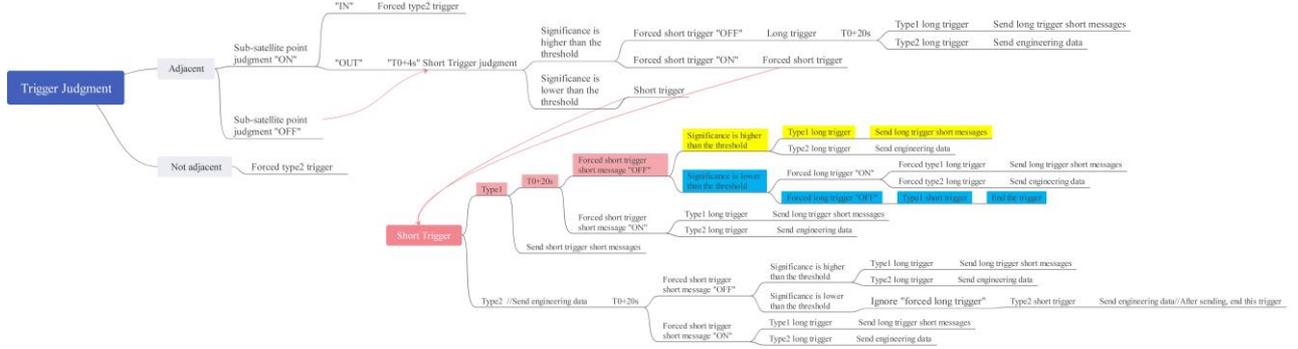

Figure 4. The workflow of the GECAM onboard trigger and localization software.

## 3 The Algorithm Principle

### 3.1 Onboard trigger

The background estimation module calculates the average single-channel counts of GRD and CPD. There are two adjustable parameters: 1) the time duration(T) chosen for the background estimation; 2) the time interval (delta T) between the background end time and the trigger time. The default values for T and delta T is 20 s and 5 s respectively, which means that data between T0-25s ~ T0-5s will be used as background estimation.

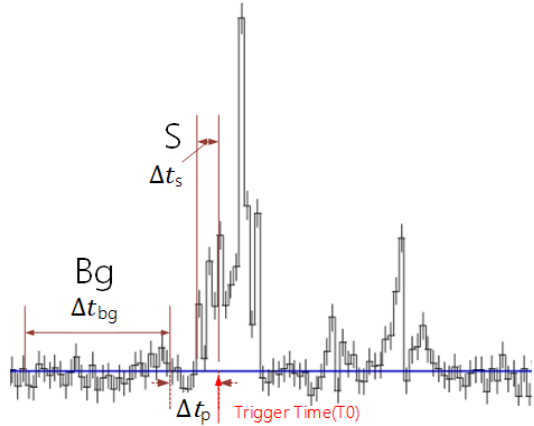

Figure 5. Illustration of the trigger algorithm.

In the trigger algorithm module, the software accumulates the total counts (denoted as S, counts in the trigger timescale) of a GRD within the trigger timescale and energy channel according to the trigger parameter table. With the background estimation (denoted as Bg, counts in the backgournd duration) provided by the background module, the Signal to Noise Ratio (SNR) of the trigger event is computed as follows:

$$SNR = (S - \frac{Bg}{\Delta t_{bg}}\Delta t_s)/\sqrt{\left(\sqrt{S}\right)^2 + \left(\frac{\sqrt{Bg}}{\Delta t_{bg}}\Delta t_s\right)^2} \quad (1)$$

Where $\Delta t_s$ denotes trigger time scale, S stands for the total counts received during $\Delta t_s$ in one GRD, $\Delta t_{bg}$ denotes the time range used for background estimation, Bg is the total counts in $\Delta t_{bg}$. As each observation is independent and obey the Poisson distribution[4], the significance of this trigger event can be calculated using Eq (1).

As shown in Table 1, the trigger time scales span from 50 ms to 4 s and the energy ranges also cover both soft band (8-25 keV) and hard band (50-400 keV).

Table 1. The trigger time scales and energy ranges.

| trigger time scale(s) | energy range(keV) |
|---|---|
| 0.05 | 15~400 |
|  | 50~400 |
|  | 15~50 |
|  | 8~25 |
| 0.1, 0.2, 0.5, 1, 2, 4 | 15~400 |
|  | 50~400 |
|  | 25~200 |
|  | 15~50 |
|  | 8~25 |

### 3.2 Onboard localization

Limited by the computing resources onboard, we adopt the detector counts distribution method to derive the location, which is also used by previous missions. We build the pre-defined positioning templates with 3 different spectra (i.e. soft, normal and hard spectra in Band function), then we calculate the chis-quare of the measured data with these model templetes, and finally get the best location based on the goodness of fit. The positioning result (θ, φ) in the payload coordinate system is transformed to the J2000 coordinate system (RA, Dec) by using the quadrature of the satellite attitude. As shown in Figure 6, the incident angle θ is the angle between S and +Z, and φ is the angle between the projection of S in the XY plane and +X.

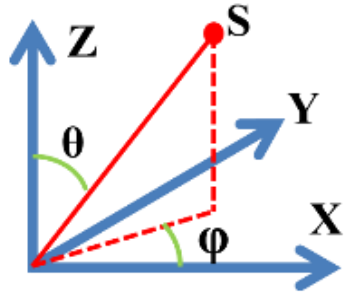

Figure 6. Definition of incident angle (θ, φ) of a celestial source (S) in the payload coordinate system.

The location template is given under the payload corrdinate system. The location template for each of the 3 spectra is a 3072 × 26 two-dimensional arrays, representing the normalized counts and normalization factors of 25 GRDs at 3072 angles. There are four location time scales in total (see Table 2). Only the first two time scales are used for short trigger, while all four time scales for long trigger. We choose the best location time scale based on the signal to noise ratio, which will be used to generate light curves and counts spectrum subsequently. The energy range for location is adjustable and the default setting is 8-400 keV.

Table 2. The time scales for location calculation.

| index | Location time scale |
|---|---|
| 1 | trigger time scale (50 ms,100 ms,200 ms,500 ms,1 s,2 s,4 s) |
| 2 | 4 s |
| 3 | 10 s |
| 4 | 17 s |

**3.3 Classification of trigger type**

Classification module aims to classify the triggered and located events. The classification for each class of sources are controlled by an adjustable table (see Table 3). By default, those events including solar flare, particle event or Earth occultation of known sources will be classfied as type II trigger, and all others as type I.

Table 3. The Classification and its type

| No. | Source class | Index | Type (adjustable) |
|---|---|---|---|
| 0 | known transient source | 7~99 | initial: Type II |
| 1 | GRB, SGR, important known transient source | 255 | initial: Type I |
| 2 | Particle event | 253 | initial: Type II |
| 3 | Earth occultation of known sources | 101~199 | initial: Type II |
| 4 | SFL | 1~5 | initial: Type II |

**4 Trigger Information**

For type I triggers, messages such as trigger time and location are downlinked to the ground through the BeiDou Navigation Satellite System (BDS) short messages. The content of the short message for short and long triggers are defined in Table 4.

Table 4. The content of BDS short messages for short and long triggers.

| No. | Trigger type | BDM index | Format code | Main infomation |
|---|---|---|---|---|
| 1 | long trigger(L) | 1-2 | 1 | Trigger, location, classification, spectrum of the highest 3 GRDs (best location time scale) |
| 2 | long trigger(L) | 3 | 2 | Light curve of the highest 3 GRDs combined |
| 3 | long trigger(L) | 4 | 3 | Light curve of the lowest 3 GRDs combined |
| 4 | long trigger(L) | 5 | 4 | Light curve of 8 CPDs combined |
| 5 | long trigger(L) | 6~30 | 5 | Light curve of GRD, one GRD per message |
| 6 | long trigger(L) | 31 | 0 | Attitude and postion of satellite at several given time points |
| 7 | short trigger(S) | 1-2 | 1 | Trigger, location, classification, spectrum of the highest 3 GRDs(best location time scale for short trigger) |
| 8 | short trigger(S) | 3~4 | 6 | Compress scheme、start time of light curve, Hight resolution light curve |
| 9 | short trigger(S) | 5~≤31 | 7 | Hight resolution light curve |

## 5 On-Ground Tests

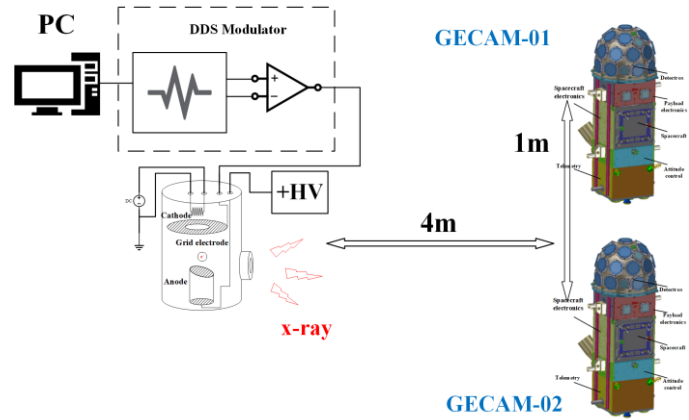

Figure 7. Illustration of the on-ground trigger test for two GECAM satellites before launch.

To test the GECAM onboard trigger and localization software, a series of trigger tests have been done with a portable x-ray tube which can mimic the emission of gamma-ray bursts. As shown in Figure 7, both GECAM satellites have been tested ([6]). During this experiment, a total of 15 types of bursts were simulated, including short, long, short and long burst etc., and all of them triggered GECAM-01 and GECAM-02 successfully.

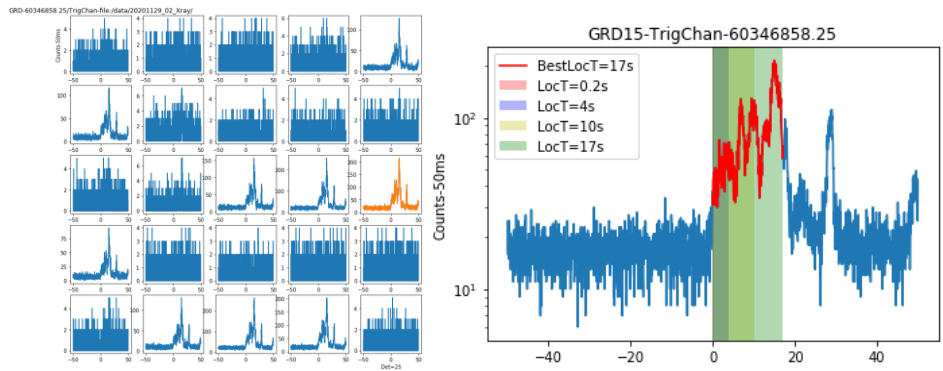

Figure 8. The light curves of the 7th trigger of GECAM-02 satellite.

Figure 8 shows the light curves of the 7th trigger of GECAM-02, the shadow regions with color indicate the location time scales, and the BestLocT given by the software is coincident with off-line analysis. In addition, all important information such as trigger time, short or long trigger judgment, classification, are consistent with expectations.

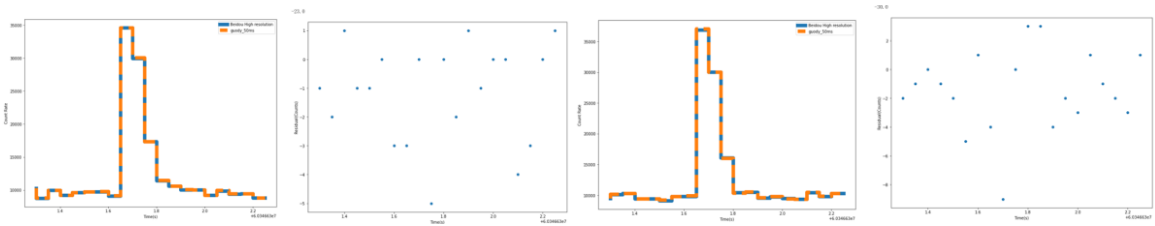

Figure 9. The high-resolution light curve of GECAM-01(left) and GECAM-02(right)

As shown in Figure 9, the high-resolution light curve recovered from BDS short messages for a short trigger is consistent with the corresponding 50 ms BTIME data.

## 6 In-flight Tests

Since the launch, GECAM has triggered both Type I and Type II triggers (such as GRBs, SGRs, SFLs and Sco-X1 Earth occultation), and both long and short bursts, confirming that the onboard trigger and location software works normally. The novel application of BDS short message to transmit astronomical alerts (trigger information) from satellite to ground was successfully demonstrated for the first time, and the time latency of the first short message for each trigger is about 1 minute. Figure 10 shows the light curve of the first long trigger received via BDS. Figure 11 shows the high-resolution light curve for the first on-orbit short trigger and its in-flight localization result.

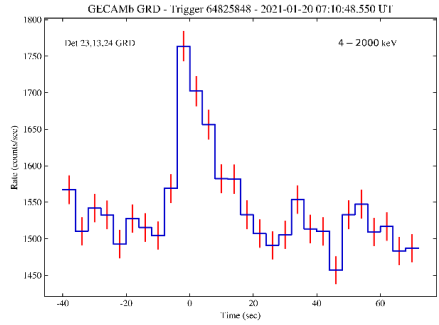

Figure 10. The first burst downlinked in real-time via BDS short message service.

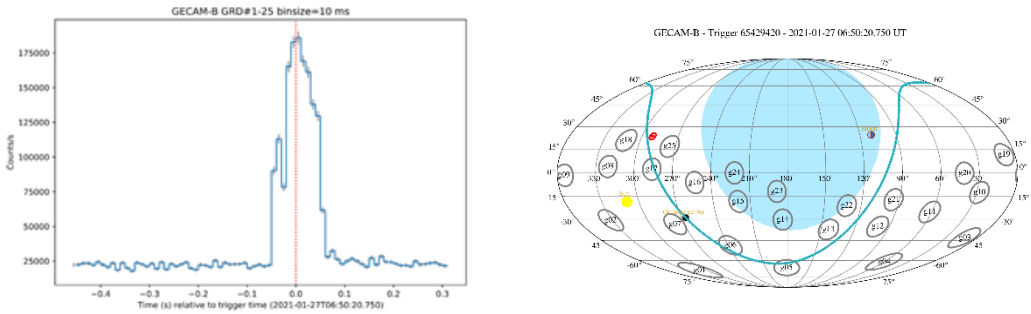

Figure 11. The high-resolution light curve (left panel) and the in-flight localization (right panel) for the first short trigger which is from SGR J1935+2154.

As shown in Table 5, the GECAM onboard trigger and localization software works smoothly and produce correct triggers for GRBs, SGRs and X-ray bursts from X-ray binaries.

## 7 Discussion and conclusion

We developed a dedicated on-board trigger and localization software running on a microcontroller Unit (CPU) of GECAM EBOX. It could not only trigger and localize various bursts, but also can classify whether the burst is long or short, and judge whether the burst is valuable (Type I trigger) or not (Type II trigger). For valueable triggers, it can generate trigger information which would be transmitted to the ground through the BDS short message service in near-real-time. The time latency of the first short messge for each trigger is about 1 minute, and all 31 short messages can be received in about 10 minutes. Based on the on-ground and in-flight tests, the performances of the onboard trigger and localization software meet the design goal. This software could be applied to other GRB missions.

Table 5. A sample of bursts triggered GECAM with this software.

| No. | GCN | title | short or long burst | trigger time (UTC) | RA (deg.) | Dec(deg.) | err(deg.) | triggerTime scale(s) | trigger energy range(keV) |
|---|---|---|---|---|---|---|---|---|---|
| 1 | GCN#29338 | GECAM In-Flight Trigger of GRB 210120A | long | 2021-01-20T07:10:48.550 | 151.37 | 55.27 | 3.19 | 4 | 15-400 |
| 2 | GCN#29347 | GRB 210121A: GECAM detection | long | 2021-01-21T18:41:48.800 | 22.12 | -49.51 | 1.25 | 0.1 | 15-400 |
| 3 | GCN#29350 | GECAM detection of a burst possibly from the X-ray burster 4U 0614+09 or GRB 210124A | long | 2021-01-24T11:50:03.600 | 102.97 | 27.2 | 3.1 | 4 | 8~25 |
| 4 | GCN#29356 | GRB 210126A: GECAM detection | long | 2021-01-26T10:00:10.600 | 106.05 | -56.26 | 6.1 | 4 | 15-400 |
| 5 | GCN#29363 | GECAM detection of a short burst probably from SGR 1935+2154 | short | 2021-01-27T06:50:20.750 | 291.31 | 23.67 | 2.16 | 0.05 | 15-400 |
| 6 | GCN#29377 | GECAM observations of SGR 1935+2154 | short | 2021-01-30T17:40:54.800 | 294.73 | 23.09 | 1 | 0.05 | 15-400 |
| 7 | GCN#29392 | GRB 210204A: GECAM detection | long | 2021-02-04T06:30:00.600 | 122.98 | 5.07 | 3.33 | 2 | 15-400 |
| 8 | GCN#29486 | GRB 210207B: GECAM detection | long | 2021-02-07T21:52:14.050 | 253.74 | 61.85 | 1 | 2 | 50~400 |
| 9 | GCN#29588 | GRB 210228A: GECAM detection | long | 2021-02-28T06:38:32.600 | 85.58 | -42.45 | 1.79 | 1 | 15-400 |
| 10 | GCN#29614 | GRB 210307B: GECAM detection | short | 2021-03-07T05:56:39.100 | 125.6 | 17.5 | 7.3 | 0.5 | 50~400 |
| 11 | GCN#29660 | GRB 210317A: GECAM detection | long | 2021-03-17T09:08:28.550 | 157.06 | -70.05 | 2.41 | 1 | 15-400 |
| 12 | GCN#29736 | GRB 210330A: GECAM detection | long | 2021-03-30T12:45:46.600 | 168.89 | -48.81 | 4.67 | 1 | 15-400 |
| 13 | GCN#29737 | GRB 210328A: GECAM detection | long | 2021-03-28T20:45:17.900 | 258.57 | -17.3 | 11.41 | 4 | 8~25 |
| 14 | GCN#29758 | GRB 210401A: GECAM detection | long | 2021-04-01T23:21:14.350 | 269.56 | -33.64 | 5.48 | 1 | 15-400 |
| 15 | GCN#29783 | GRB 210409A: GECAM detection | long | 2021-04-09T21:28:07.950 | 69.4 | -59.31 | 2.77 | 0.5 | 15-400 |
| 16 | GCN#29813 | GRB 210413A: GECAM detection | long | 2021-04-13T01:07:25.600 | 68.6 | 11.3 | 3.6 | 2 | 15-400 |
| 17 | GCN#29899 | GRB 210425A: GECAM detection | short | 2021-04-25T07:07:04.200 | 67.53 | -51.76 | 2.73 | 0.2 | 15-400 |
| 18 | GCN#29906 | GRB 210427A: GECAM detection | long | 2021-04-27T04:57:13.100 | 175.26 | -59.87 | 5.75 | 0.2 | 30-800 |
| 19 | GCN#29990 | GRB 210511B: GECAM detection | long | 2021-05-11T11:26:40.600 | 317.99 | 59.53 | 3.19 | 0.5 | 30-800 |

**Acknowledgements**

The GECAM (Huairou-1) mission is supported by the Strategic Priority Research Program on Space Science, the Chinese Academy of Sciences, Grant No. XDA15360000. The authors thank supports from the GECAM satellite operation and support teams and the Beidou Satellite Navigation System.